\documentstyle[prl,aps,epsfig,twocolumn]{revtex}

\catcode`\@=11

\def\maketitle2{\par 
\begingroup
\let\cite\@bylinecite
\def\thefootnote{\fnsymbol{footnote}}%
\twocolumn[\@maketitle2\vskip2pc]%
\thispagestyle{plain}\@thanks
\endgroup
\def\thefootnote{\arabic{footnote}}%
\setcounter{footnote}{0}%
\let\maketitle2\relax \let\@maketitle2\relax
\let\@thanks\relax \let\@authoraddress\relax \let\@title\relax
\let\@date\relax \let\thanks\relax \let\@abstract\relax 
\let\@pacs\relax}

\def\abstract#1{\gdef\@abstract{{\par 
\bgroup
\ifdim\prevdepth=-1000pt \prevdepth0pt\fi
\hsize\columnwidth
\dimen0=-\prevdepth \advance\dimen0 by17.5pt \nointerlineskip
\small\vrule width 0pt height\dimen0 \relax}{~~}#1\egroup}}

\def\pacs#1{\gdef\@pacs{{\par 
\bgroup
\hsize\columnwidth \parindent0pt
\ifdim\prevdepth=-1000pt \prevdepth0pt\fi
\dimen0=-\prevdepth \advance\dimen0 by20pt\nointerlineskip
\egroup} PACS numbers:~#1}}

\def\@maketitle2{
\@preprint
\@title
\ifdim\prevdepth=-1000pt \prevdepth0pt\fi
\@authoraddress
\@date
\begin{list}{}{\leftmargin=0.10753\textwidth \rightmargin=\leftmargin
\itemsep=1pc\partopsep=-1pc}
\item\@abstract
\item\@pacs
\end{list}
}

\catcode`\@=12

\begin{document}
%
%
\title{Microscopic calculation of the inclusive electron scattering 
       structure function in $^{16}$O}
\author{Bogdan~Mihaila\thanks{electronic mail:Bogdan.Mihaila@unh.edu}}
\address{
Department of Physics, University of New Hampshire, Durham, NH 03824 \\
Physics Division, Oak Ridge National Laboratory, P.O. Box 2008,
Oak Ridge, TN 37831}
\author{Jochen~H.~Heisenberg\thanks{electronic mail:Jochen.Heisenberg@unh.edu} }
\address{Department of Physics, University of New Hampshire,
Durham, NH 03824}

\date{\today}

\abstract{ We calculate the charge form factor and the longitudinal
  structure function for $^{16}$O and compare with the available
  experimental data, up to a momentum transfer of 4 fm$^{-1}$.  The
  ground state correlations are generated using the coupled cluster
  [$\exp({\bf S})$] method, together with the realistic v$18$ $NN$
  interaction and the Urbana IX three-nucleon interaction.
  Center-of-mass corrections are dealt with by adding a center-of-mass
  Hamiltonian to the usual internal Hamiltonian, and by means of a
  many-body expansion for the computation of the observables measured
  in the center-of-mass system.  }

\maketitle2 


One of the fundamental problems in nuclear physics is related to
developing a complete understanding of how nuclear structure arises as
a result of the underlying interaction between nucleons. This in turn
should help us develop a complete understanding of the electromagnetic
structure of the nucleus, as revealed by the wealth of high-quality
data that electron scattering experiments have provided for the past
30 years.  The interplay of nuclear correlations, meson-exchange
current or charge densities, relativistic effects in nuclei, the
importance of three- or more many-body interactions in relation with
the dominant two-body interaction in nuclei awaits to being assessed
in greater detail.  Unfortunately, solutions of the many-body
Schr\"odinger equation with realistic interactions have been proven
very difficult to obtain. Only in recent years has progress been made
and first results of microscopic calculations relating to ground-state
and low-excited states for nuclei with $A\le 7$ been
reported~\cite{ref:Pudliner_etal}.  These calculations have been
obtained using the Green's function Monte Carlo method, but this
approach, just like the Fadeev~\cite{ref:Fadeev} or the Correlated
Hyperspherical Harmonics~\cite{ref:CHH} methods successfully used for
the $A=3,4$-body system, suffers limitations in the number of nucleons
they can treat.  To date, only the Variational Monte Carlo
method~\cite{ref:Pieper_etal} has enjoyed success in solving the
many-body problem for medium nuclei, but those results still show a
room for improvement.

We are using the $exp({\bf S})$ coupled-cluster expansion to calculate
the ground state of $^{16}$O.  Our approach is very similar to the
standard approach, first developed by the Bochum
group~\cite{ref:Kummel_etal}, and has been outlined recently
in~\cite{ref:paper_one}.  The idea behind the coupled-cluster
expansion formalism relies on the ability of expanding the model
nuclear wave function in the many-body Hilbert space in terms of two
Abelian subalgebras of multiconfigurational creation and their
Hermitian-adjoint destruction operators.  The expansion coefficients
carry then the interpretation of nuclear correlations.  The fact that
we make no artificial separation between ``short-range'' and
``long-range'' correlations is one particular strength of this
many-body method.

The derivation of the explicit equations is quite tedious, but
requires only standard techniques.  For a closed-shell nuclear system,
the total Hamiltonian is given as
\begin{equation}
   {\bf H}
   \ = \
   \sum_i T_i
   \ + \ \sum_{i<j} \ V_{ij}
   \ + \ \sum_{i<j<k} \ V^{tni}_{ijk}
   \>.
\label{eq:Hamiltonian_0}
\end{equation}
The Hamiltonian includes a nonrelativistic one-body kinetic energy, a
two-nucleon potential, and a supplemental three-nucleon potential.  We
have chosen the Argonne $v$18 potential~\cite{ref:argonne_v18} as the
most realistic nucleon-nucleon interaction available today.  The
Argonne $v$18 model provides an accurate fit for both $pp$ and $nn$
scattering data up to 350 MeV with a $\chi^2$/datum near one.  The
introduction of charge-independence breaking in the strong force is
the key element of obtaining this high performance.  However, the
two-body part of this interaction results in over-binding and too
large a saturation density in nuclear matter. Therefore, the $NN$
potential is supplemented by a three-nucleon interaction (part of the
Urbana family~\cite{ref:tnipot}), which includes a long-range two-pion
exchange and a short-range phenomenological component.  The Urbana-IX
potential is adjusted to reproduce the binding energy of $^3$H and
give reasonable saturation density in nuclear matter when used with
Argonne $v$18~\cite{ref:Pudliner_etal}.

We are searching for the correlated ground state of the Hamiltonian
$H$, which we denote by $| \tilde 0 \rangle$.  The ansatz for the
many-body wave function $| \tilde 0 \rangle$ is defined as the result
of the cluster correlation operator, ${\rm S}^{\dag}$, acting on the
reference state of the many-body system, the uncorrelated ground
state~$|0\rangle$:
\[
   | \tilde{0} \rangle \ = \ {\displaystyle e^{{\bf S}^{\dag}}} | 0 \rangle
   \>.
\]
For a number-conserving Fermi system, the standard choice for $| 0
\rangle $ is the single-particle shell-model (Slater determinant)
state formed from an antisymmetrized product of single-particle wave
functions.  The cluster correlation operator is defined in terms of
its {\em ph}-creation operators expansion (${\bf O}^{\dag}_0$~=~${\bf
  1}$, ${\bf O}^{\dag}_1$~=~${\bf a}^{\dag}_{p_1} {\bf a}_{h_1}$,
${\bf O}^{\dag}_2$~=~${\bf a}^{\dag}_{p_1} {\bf a}^{\dag}_{p_2} {\bf
  a}_{h_2} {\bf a}_{h_1}$) as
\[
   {\bf S}^{\dag} = \sum_{n=0}^\infty \frac{1}{n!} S_n {\bf O}^{\dag}_n
   \>.
\]
The problem of solving for the many-body wave function $| \tilde 0
\rangle$ and the ground-state energy, $E$, is now reduced to the
problem of solving for the amplitudes $S_n$.  This implies solving a
set of non-linear equations, which may be obtained using a variational
principle.  We construct a variation $\delta |\tilde 0 \rangle$
orthogonal to the correlated ground state as
\[
   \delta \, | \tilde 0 \rangle
   \ = \
   e^{- \, {\bf S}} \ {\bf O}^{\dag}_n \
   | 0 \rangle
   \>,
\]
and require that the Hamiltonian between the ground state and such a 
variation vanishes. As a result, we obtain an equation for the
ground-state energy eigenvalue $E$ in terms of the cluster
correlation coefficients, $\{ S_n \}$, 
and a set of formally exact coupled nonlinear equations for 
these coefficients:
\begin{eqnarray*}
   E & = & \langle \tilde 0 | \ e^{\bf S} \ {\bf H} \ e^{- \, {\bf S}} \ 
           | \tilde 0 \rangle
   \>,
   \\
   0 & = &  
   \langle 0 | \ e^{\bf S} \ {\bf H} \ e^{- \, {\bf S}} \ 
   {\bf O}^{\dag}_n \
   | 0 \rangle
   \>.
\end{eqnarray*}
Then, the computation breaks down into two steps~\cite{ref:paper_one}:
In the first step, the G-matrix interaction is calculated inside the
nucleus including all the corrections.  This results in amplitudes for
the 2$p$2$h$ correlations, which are implicitly corrected for the
presence of 3$p$3$h$ and 4$p$4$h$ correlations.  In the second step
the mean field is calculated from these correlations and the
single-particle Hamiltonian is solved to give mean-field
eigenfunctions and single-particle energies.  These two steps are
iterated until a stable solution is obtained.  Calculations are
carried out entirely in configuration space where a $50\hbar\omega$
space is used.  The general approach, when the Hamiltonian includes
only up to two-body operators, has been presented
in~\cite{ref:paper_one,ref:terms}.  The results we report here have
been obtained by taking into account the three-nucleon interaction via
a density-dependent approach.  The details of this approach will be
presented elsewhere~\cite{ref:paper_part2}.

Once the correlated ground state $| \tilde 0 \rangle$ is obtained, 
we can calculate the expectation value of any arbitrary operator $A$ as
\[
   \bar{a} =
   \langle 0 \, | \,
             e^{\bf S} \, A \, e^{- \bf S} \, \tilde{\bf S}^{\dag}
             \, | \, 0 \rangle
   \>,
\]
where $\tilde{\bf S}^{\dag}$ is also defined by
its decomposition in terms of {\em ph}-creation operators
\[
   \tilde{\bf S}^{\dag} =
   \sum_n \frac{1}{n!} \tilde{S}_n {\bf O}^{\dag}_n
   \>.
\]
The amplitudes $\tilde{S}_n$ are obtained in terms of the ${S}_n$
amplitudes in an iterative fashion.

Note that the correlated ground state $| \tilde 0 \rangle$ is not
translationally invariant since it depends on the $3A$ coordinates of
the nucleons in the laboratory frame.  Therefore, in practice one has
to take special care of correcting for the effects of the
center-of-mass motion. This is done in several steps: First, the
Hamiltonian~(\ref{eq:Hamiltonian_0}) is replaced by the {\em internal}
Hamiltonian
\[
   H_{int} 
   \ = \ H \ - \ T_{CM}
   \>,
\]
which is now entirely written in the center-of-mass frame by removing
the center-of-mass kinetic energy, $T_{CM} = P_{CM}^2/(2 \, m \, A)$,
with $m$ the nucleon mass.  Both the two- and three-nucleon
interactions are given in terms of the relative distances between
nucleons, so in this respect no corrections are needed.
%
%
Secondly, a many-body expansion has been devised~\cite{ref:paper_two} 
in order to carry out the necessary corrections required by the 
calculation of observables, which are measured experimentally 
in the center-of-mass frame. 
This procedure is based on the assumption that we can neglect the
correlations between the center-of-mass and relative coordinates
degrees of freedom, and a factorization of the correlated ground state
$| \tilde 0 \rangle$ into components which depend only on the center-of-mass
and the relative coordinates, respectively, is possible.  We also
assume that, indeed, the correlated ground state $| \tilde 0 \rangle$
provides a good description of the internal structure of the nucleus.
Finally, in order to ensure such a separation, a supplemental
center-of-mass Hamiltonian is added
\[
  H_{CM} \ = \ 
  \beta_{CM} \, 
   \left [ T_{CM} \ + \ \frac{1}{2} \ (m \, A) \ \Omega^2 \ R_{CM}^2
   \right ]
  \>,
\]   
which has the role of constraining the center-of-mass component 
of the ground-state wave function~\cite{ref:CM_people}.
We choose the values of the parameters $\beta_{CM}$ and $\Omega^2$
such that they correspond to a value domain for which the binding
energy
\[
E \ = \ \langle H_{int}^{'} = H_{int} + H_{CM} \rangle 
            \, - \, \langle H_{CM} \rangle
\]             
is relatively insensitive to the choice of the $\beta_{CM}$ and
$\Omega^2$ values~\cite{ref:paper_part2}.  When leaving out the
center-of-mass Hamiltonian, the calculated binding energy of $^{16}$O
is equal to 7.54 Mev/nucleon, which is thought as a reasonable value,
given the uncertainties related to the three-nucleon interaction.

\begin{figure}[h]
   \epsfxsize = 3.0in
   \centerline{\epsfbox{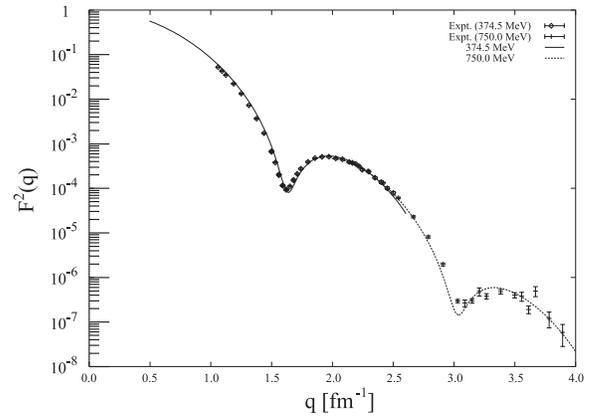}}
   \caption{Charge form factor for O-16,
            obtained in the DWBA picture ($v_{18}$ + UIX + ME)}
\label{fig:o16_elastic}
\end{figure}

Figure~\ref{fig:o16_elastic} shows the theoretical
result for the charge form factor in $^{16}$O.  In the one-body
Born-approximation picture, the charge form factor is given as
\[
   F_{L}(q) \ = \
   \langle
   \tilde 0 \, | \,
          \sum_k \, f_k(q^2) \ e^{i \vec{q} \cdot \vec r_k'}
          \, | \, \tilde 0 \rangle
   \>,
\]
with $f_k(q^2)$ the nucleon form factor, which takes into account the
finite size of the nucleon $k$~\cite{ref:ffn_Iach}.  We also take into
account the {\em model-independent} part of the ``Helsinki
meson-exchange model"~\cite{ref:ffn_disc}, namely the contributions
from the $\pi$- and $\rho$-exchange ``seagull" diagrams, with the
pion- and $\rho$-meson propagators replaced by the Fourier transforms
of the isospin-dependent spin-spin and tensor components of the $v$18
$NN$ interaction. This substitution is required in order for the
exchange current operator to satisfy the continuity equation together
with the interaction model.  The contributions of the $\pi$- and
$\rho$-exchange charge density give a measurable correction only for
$q>2$ fm$^{-1}$.

In order to generate the form factor depicted in
Fig.~\ref{fig:o16_elastic}, we have first used the procedures
of~\cite{ref:paper_two}, keeping the contributions that can be written
in terms of the one- and two-body densities. Then we take the Fourier
transform in order to produce the theoretical charge density. Using
this theoretical charge density, we generate the charge form factor in
a Distorted Wave Born Approximation picture~\cite{ref:DWBA}, in order
to take into account the distortions due to the interaction of the
electron with the Coulomb field.  This last step results in smoothing
out the sharp diffraction minima usually seen in the calculated charge
form factor~\cite{ref:Pieper_etal,ref:paper_two}.  The agreement with
the experiment is reasonably good over the whole range of $q$ spanned
by the available experimental data~\cite{ref:SickMcCarthy}.

A second electron scattering observable that we would like to compare
with is the longitudinal structure function $S_L(q)$, sometimes
called the Coulomb sum rule, which is sensitive to the short-range
correlations induced by the repulsive core of the $NN$
interaction~\cite{ref:SL_ppdf,ref:Schiavilla_Carlson}.  The Coulomb
sum rule, $S_L(q)$, represents the total integrated strength of the
longitudinal response function measured in inclusive electron
scattering.  In the nonrelativistic
limit~\cite{ref:Schiavilla_Carlson}, we have
\[
   S_L(q)
   \ \equiv \ 
   1 + \rho_{LL}(q) 
      -
   \frac{1}{Z} | \langle \tilde 0 | \rho(q) | \tilde 0 \rangle |^2
   \>,
\]
where 
$\rho(q)$ is the nuclear charge operator
\[
   \rho(q)
   \ = \ 
   \frac{1}{2} \ 
   \sum_i^A \, e^{i \vec q \cdot \vec r_i} \ ( 1 + \tau_{z,\, i} )
   \>,
\]
and $\rho_{LL}(q)$ is the longitudinal-longitudinal distribution function 
\[
   \rho_{LL}(q)
   \ = \
   \int d \vec r_1 \int d \vec r_2 \ 
   j_0(q |\vec r_1 - \vec r_2|) \ \rho^{(p,p)}(\vec r_1,\vec r_2)
   \>.
\]
Here $\rho^{(p,p)}(\vec r_1,\vec r_2)$ is the proton-proton two-body density 
\begin{eqnarray*}
   &&
   \rho^{(p,p)}(\vec r_1,\vec r_2)
   \\ &&
   \ = \
   \frac{1}{4}
   \sum_{i,j} 
   \langle \tilde 0 | \, 
               \delta(\vec r_1 - \vec r_i) \delta(\vec r_2 - \vec r_j) 
               ( 1 + \tau_{z,\, i} ) ( 1 + \tau_{z,\, j} ) \, 
   | \tilde 0 \rangle
   \>,
\end{eqnarray*}
normalized as
\[
   \int d \vec r_1 \int d \vec r_2 \ 
   \rho^{(p,p)}(\vec r_1,\vec r_2) \ = \ Z - 1
   \>.
\]

\begin{figure}[h]
   \epsfxsize = 3.0in
   \centerline{\epsfbox{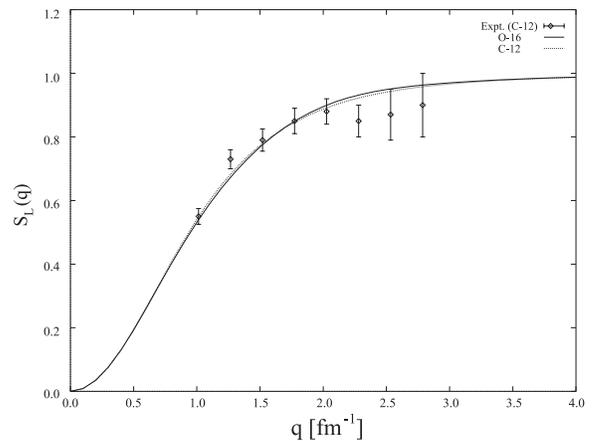}}
   \caption{Coulomb sum for C-12 and O-16,
            compared with ``experimental" C-12 data which include
            theoretically determined high-energy tail corrections}
\label{fig:sl}
\end{figure}

In light nuclei reasonable agreement between theory and experiment is
obtained for the Coulomb sum rule~\cite{ref:SL_light}.  In heavier
nuclei, however, the experimental situation is a lot more
controversial, since both a certain lack of strength has been reported
and because of the inherent difficulty of separating the longitudinal
and transverse contributions in the cross section due to the
distortion effects of the electron waves in the nuclear Coulomb field.
Figure~\ref{fig:sl} shows the calculated Coulomb sum in $^{16}$O.
Since no experimental data are available for $^{16}$O, we compare the
results of the present calculation with the $^{12}$C experimental data
from~\cite{ref:SL_c12} with an estimate~\cite{ref:SL_tail} for
contributions from large $\omega$.  The large error bars on the
experimental data are largely due to systematic uncertainties
associated with tail contribution~\cite{ref:SL_2body}.  Preliminary
theoretical results for $^{12}$C are also shown and appear to follow
closely the results for theoretical curve for $^{16}$O.

This calculation represents the most detailed calculation available
today, using the coupled cluster expansion, for a nuclear system with
$A>8$.  This also represents a contribution to the on-going effort of
carrying out microscopic calculations that directly produce nuclear
shell structure from realistic nuclear interactions.  Similar
calculations for other closed-shell nuclei in the $p$- and $sd$-shell
are currently under way.

This work was supported in part by the U.S. Department of Energy
(DE-FG02-87ER-40371).  The work of B.M. was also supported in part by
the U.S. Department of Energy under contract number DE-FG05-87ER40361
(Joint Institute for Heavy Ion Research), and DE-AC05-96OR22464 with
Lockheed Martin Energy Research Corp.  (Oak Ridge National
Laboratory).  The calculations were carried out on a dual-processor
500 MHz Pentium II PC in the Nuclear Physics Group at the University
of New Hampshire.  Some of the preliminary calculations have also run
on a 180 MHz R10000 Silicon Graphics Workstation in the Computational
and Theoretical Physics Section at ORNL.  The authors gratefully
acknowledge useful conversations with John Dawson and David Dean.

%
%

%
%

\newpage

\end{document}